\documentclass[%
preprint,
superscriptaddress,
 amsmath,amssymb,
 aps,
prb,
]{revtex4-1}

\usepackage{graphicx}
\usepackage{color}
\usepackage{dcolumn}
\usepackage{bm}
\usepackage{threeparttable}
\usepackage[top=20truemm,bottom=20truemm,left=20truemm,right=20truemm]{geometry}

\begin{document}
\emergencystretch 3em


\title{Electronic structure of the novel high-$T_{\rm C}$ ferromagnetic semiconductor (Ga,Fe)Sb: x-ray magnetic circular dichroism and resonance photoemission spectroscopy studies}

\author{S. Sakamoto}
\affiliation{Department of Physics, The University of Tokyo, Bunkyo-ku, Tokyo 113-0033, Japan}

\author{N. T. Tu}
\affiliation{Department of Electrical Engineering and Information Systems, The University of Tokyo, Bunkyo-ku, Tokyo 113-8656, Japan}

\author{Y. Takeda}
\affiliation{Materials Sciences Research Center, Japan Atomic Energy Agency (JAEA), Sayo-gun, Hyogo 679-5148, Japan}

\author{S.-i. Fujimori}
\affiliation{Materials Sciences Research Center, Japan Atomic Energy Agency (JAEA), Sayo-gun, Hyogo 679-5148, Japan}

\author{P. N. Hai}
\affiliation{Department of Physical Electronics, Tokyo Institute of Technology, Meguro-ku, Tokyo 152-0033, Japan}

\author{L. D. Anh}
\affiliation{Department of Electrical Engineering and Information Systems, The University of Tokyo, Bunkyo-ku, Tokyo 113-8656, Japan}

\author{Y. K. Wakabayashi}
\affiliation{Department of Electrical Engineering and Information Systems, The University of Tokyo, Bunkyo-ku, Tokyo 113-8656, Japan}

\author{G. Shibata}
\affiliation{Department of Physics, The University of Tokyo, Bunkyo-ku, Tokyo 113-0033, Japan}

\author{M. Horio}
\affiliation{Department of Physics, The University of Tokyo, Bunkyo-ku, Tokyo 113-0033, Japan}

\author{K. Ikeda}
\affiliation{Department of Physics, The University of Tokyo, Bunkyo-ku, Tokyo 113-0033, Japan}

\author{Y. Saitoh}
\affiliation{Materials Sciences Research Center, Japan Atomic Energy Agency (JAEA), Sayo-gun, Hyogo 679-5148, Japan}

\author{H. Yamagami}
\affiliation{Materials Sciences Research Center, Japan Atomic Energy Agency (JAEA), Sayo-gun, Hyogo 679-5148, Japan}
\affiliation{Department of Physics, Kyoto Sangyo University, Kyoto 603-8555, Japan}

\author{M. Tanaka}
\affiliation{Department of Electrical Engineering and Information Systems, The University of Tokyo, Bunkyo-ku, Tokyo 113-8656, Japan}
\affiliation{Center for Spintronics Research Network, The University of Tokyo,  Bunkyo-ku, Tokyo 113-8656, Japan}

\author{A. Fujimori}
\affiliation{Department of Physics, The University of Tokyo, Bunkyo-ku, Tokyo 113-0033, Japan}

\date{\today}

\begin{abstract}
The electronic structure and the magnetism of the novel ferromagnetic semiconductor (Ga,Fe)Sb, whose Curie temperature $T_{\rm C}$ can exceed room temperature, were investigated by means of x-ray absorption spectroscopy (XAS), x-ray magnetic circular dichroism (XMCD), and resonance photoemission spectroscopy (RPES).
The line-shape analyses of the XAS and XMCD spectra suggest that the ferromagnetism is of intrinsic origin.
The orbital magnetic moments deduced using XMCD sum rules were found to be large, indicating that there is a considerable amount of 3$d^{6}$ contribution to the ground state of Fe.
From RPES, we observed a strong dispersive Auger peak and non-dispersive resonantly enhanced peaks in the valence-band spectra. The latter is a fingerprint of the correlated nature of Fe 3$d$ electrons, whereas the former indicates their itinerant nature.
It was also found that the Fe 3$d$ states have finite contribution to the DOS at the Fermi energy. These states presumably consisting of majority-spin $p$-$d$ hybridized states or minority-spin $e$ states would be responsible for the ferromagnetic order in this material.
\end{abstract}

\pacs{Valid PACS appear here}
\maketitle


\section{Introduction}

To create functional devices exploiting the spin degree of freedom in semiconductors has been one of the major challenges in the field of electronics \cite{Wolf:2001aa, Igor, Awschalom:2007aa}. 
Under such circumstances, magnetically doped semiconductors, or diluted magnetic semiconductors (DMSs), have attracted much attention since they possess both magnetic and semiconducting properties \cite{Furdyna:1988aa, Jungwirth:2006aa, Sato:2010aa, Dietl:2014aa, Jungwirth:2014aa, Tanaka:2014aa}.  Mn-doped III-V semiconductors such as (In,Mn)As \cite{Munekata:1989aa, Ohno:1992aa} and (Ga,Mn)As \cite{Ohno:1996aa, Hayashi:1997aa, Van-Esch:1997aa} have been extensively studied because they exhibit carrier-induced ferromagnetism, where the ferromagnetic interaction between the Mn magnetic moments is mediated by hole carriers, and it is possible to control the ferromagnetism through changing the carrier concentration by gate voltage \cite{ohno2000electric, Sawicki:2010aa} or light irradiation \cite{Koshihara:1997aa}.
Despite those attractive features, they have also shortcomings for practical applications: their Curie temperatures ($T_{\rm C}$) are much lower than room temperature, 90 K for (In,Mn)As \cite{Schallenberg:2006aa} and 200 K for (Ga,Mn)As \cite{Chen:2011aa}; only $p$-type conductivity is realized since Mn always acts as an acceptor at the substitutional In$^{3+}$ or Ga$^{3+}$ sites.

Recently Fe-doped ferromagnetic III-V semiconductors (In,Fe)As:Be \cite{Nam-Hai:2012aa, Nam-Hai:2012ab, Nam-Hai:2012ac}, (Ga,Fe)Sb \cite{Tu:2014aa, Tu:2015aa, Tu:2016aa}, (Al,Fe)Sb \cite{Anh:2015aa}, and (In,Fe)Sb \cite{Tu:2018aa, Tu:2018ab, Kudrin:2017aa} were synthesized, and exhibit some advantages over the Mn-doped ones.
If Fe substitutes for the In$^{3+}$ or Ga$^{3+}$ site and takes the stable valence of 3+ with the 3$d^{5}$(4$sp$)$^{3}$ configuration, no charge carrier will be provided and hence both $n$- and $p$-type conduction will be possible via additional carrier doping. 
In fact, (Al,Fe)Sb is insulating, (In,Fe)As:Be $n$-type, where doped interstitial Be atoms act as double donors, and (Ga,Fe)Sb $p$-type, where native charged defects such as Ga anti-sites are thought to act as acceptors and provide holes.
The $T_{\rm C}$'s of these materials are relatively high, 70 K for (In,Fe)As:Be \cite{Nam-Hai:2012ac}, 40 K for (Al,Fe)Sb \cite{Anh:2015aa}, 335 K for (In,Fe)Sb \cite{Tu:2018aa}, and 340 K for (Ga,Fe)Sb \cite{Tu:2016aa}. 
It has been found that the distribution of Fe atoms is non-uniform in the zinc-blende crystal structure of these materials, which seems to play an important role in stabilizing the ferromagnetic order \cite{Sakamoto:2016aa, Tu:2015aa, Anh:2015aa}.
However, the microscopic origin of the ferromagnetism in terms of their electronic structures has not been clarified yet and remains to be investigated. For this purpose, we have performed soft x-ray absorption spectroscopy (XAS), x-ray magnetic circular dichroism (XMCD), and resonance photoemission spectroscopy (RPES) studies of (Ga,Fe)Sb.

XAS and XMCD at the $L_{2,3}$ absorption edges of the 3$d$ transition metals are very powerful methods for the purpose of clarifying the electronic structures related to the ferromagnetism. 
Since x-ray absorption takes place at a specific constituent atom, one can obtain element-specific information about the electronic structure and its relation to the magnetism, excluding extrinsic effects such as diamagnetic contribution from the substrate.
XMCD sum rules make it possible to obtain the spin and orbital magnetic moments of the constituent atoms separately \cite{Carra:1993aa,Thole:1992aa}. 
In addition to XAS and XMCD, resonance photoemission spectroscopy (RPES) has been frequently employed as a direct probe to examine the electronic structure of materials.
RPES provides the information about the partial density of states (PDOS) of 3$d$ transition-metal element, and has been used to study the electronic structures of ferromagnetic semiconductors (FMSs) such as (Ga,Mn)As \cite{Rader:2004aa, Kobayashi:2014aa}, Ge:Fe \cite{Sakamoto:2017aa}, and (Ba,K)(Zn,Mn)$_{2}$As$_{2}$ \cite{Suzuki:2015aa, Suzuki:2015ab}. 
Moreover, the combination of RPES and XMCD yields the PDOS of only ferromagnetically active components, and is suitable for studying FMSs, where doped magnetic atoms are often oxidized at the surface.

\section{Experiment}

Ga$_{1-x}$Fe$_{x}$Sb films with two different Fe contents $x= 6.0$ and $13.7 \%$ (referred to as sample A and B, respectively) were grown on GaAs(001) substrates using the low-temperature molecular beam epitaxy (LT-MBE) method. To relax the lattice mismatch between (Ga,Fe)Sb and GaAs, three buffer layers were inserted; initially GaAs (50 nm) and AlAs (10 nm) layers were successively grown at the substrate temperature $T_{S}$ of 550 $^{\circ}$C, and then an AlSb (100 nm) layer at $T_{S}=470 ^{\circ}$C.
After growing the buffer layers, the (Ga,Fe)Sb layer of 50 nm thickness was grown. Here, the $T_{S}$ was set to 200 $^{\circ}$C for sample A and 250 $^{\circ}$C for sample B.
Lastly sub-nanometer-thick amorphous As cap layer was deposited to prevent surface oxidation.
Note that sample A was paramagnetic down to 5 K, and sample B was ferromagnetic with $T_{\rm C}$ = 170 K. 
In order to remove the oxidized surface, we etched the sample by hydrochloric acid (HCl) (2.4 mol/L) for 5 seconds and subsequently rinsed it by water just before loading the sample in the vacuum chamber of the spectrometer \cite{Edmonds:2004aa, Sakamoto:2017aa}. 

All the measurements were performed at beam line BL23SU of SPring-8. 
Circularly polarized x rays of 690 - 780 eV were used for both absorption and photoemission measurements.
For XMCD measurements, a magnetic field was applied parallel to the incident x rays and perpendicular to the sample surface.
Absorption signals were taken in the total electron yield (TEY) mode, and dichroic signals were measured by reversing the helicity of x rays with 1 Hz frequency at each photon energy under a fixed magnetic field. 
In order to eliminate spurious XMCD signals, the scans were repeated with opposite magnetic field directions. Each XMCD spectrum was obtained as the average, 
namely, ($(\sigma_{+,h} - \sigma_{-,h}) +  (\sigma_{-,-h} - \sigma_{+,-h}))$, where $\sigma$ denotes the absorption cross-sections, the first subscript the helicity of x rays, and the second subscript the sign of the magnetic field. XAS was obtained as the summation of all the four terms.
Note that two-step inverse tangent function representing the Fe $L_{2,3}$-edge jumps were subtracted from each term \cite{Chen:1995aa}.

For RPES measurements, the sample temperature was set to 100 K, and the energy resolution was about 150 meV. 
The samples were placed so that the [-110] direction became parallel to the analyzer slit and perpendicular to the beam. Photoelectrons were collected in the normal emission geometry with 45-degree light incidence.

\section{Results and Discussion}
\subsection{XAS and XMCD}

\begin{figure}
\begin{center}
\includegraphics[width=8.5 cm]{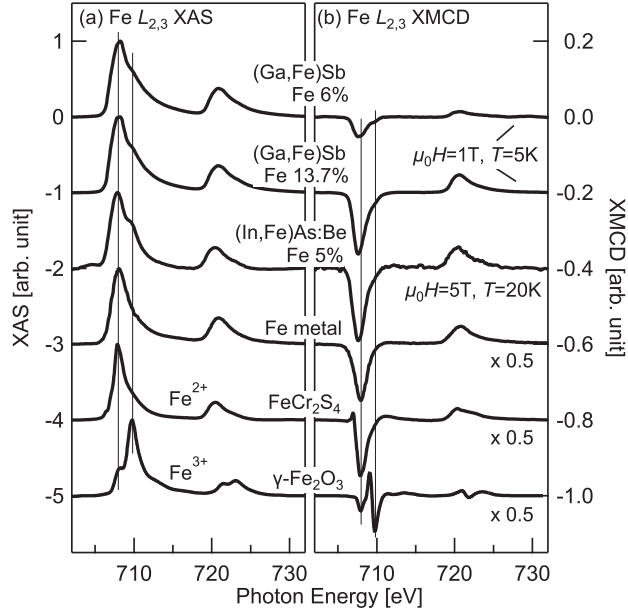}
\caption{XAS (a) and XMCD (b) spectra of (Ga,Fe)Sb compared with those of (In,Fe)As:Be \cite{Sakamoto:2016aa}, bcc Fe (Ref. 37)\cite{Chen:1995aa}, FeCr$_{2}$O$_{4}$ \cite{Verma_FeCr2S4}, and $\gamma$-Fe$_{2}$O$_{3}$ \cite{Brice-Profeta:2005aa}. The XMCD spectra of bcc Fe, FeCr$_{2}$O$_{4}$, and $\gamma$-Fe$_{2}$O$_{3}$ are multiplied by 0.5 for ease of comparison.}
\label{XMCD}
\end{center}
\end{figure}

Figures \ref{XMCD}(a) and \ref{XMCD}(b) show XAS and XMCD spectra of the present (Ga,Fe)Sb films after the HCl etching compared with those of (In,Fe)As:Be \cite{Sakamoto:2016aa}, Fe metal \cite{Chen:1995aa}, FeCr$_{2}$S$_{4}$ \cite{Verma_FeCr2S4}, and $\gamma$-Fe$_{2}$O$_{3}$ \cite{Brice-Profeta:2005aa}. 
Note that the Fe 3$d$ electrons in FeCr$_{2}$S$_{4}$ and $\gamma$-Fe$_{2}$O$_{3}$ are localized, and that the valence of Fe is 2+ in FeCr$_{2}$S$_{4}$ and 3+ in $\gamma$-Fe$_{2}$O$_{3}$. 

The line shapes of the spectra of the two (Ga,Fe)Sb films resemble those of bcc Fe metal rather than the sharper spectra of FeCr$_{2}$S$_{4}$, manifesting the itinerant nature of the Fe 3$d$ electrons in (Ga,Fe)Sb.
Although the spectral line shapes of the two (Ga,Fe)Sb films look almost identical, representing nearly the same electronic structure, the intensity of the XMCD signals was significantly suppressed for the 6\% Fe-doped sample compared with the 13.7\% Fe-doped sample simply because only the 13.7\% Fe-doped sample exhibits ferromagnetism while the 6\% Fe-doped one is paramagnetic at 5 K. 
Such an insensitivity of the line shape to transition-metal content was also reported for (Ga,Mn)As \cite{Wu:2005aa}, (In,Fe)As:Be \cite{Sakamoto:2016aa}, etc. 

\begin{figure}
\begin{center}
\includegraphics[width=7.5 cm]{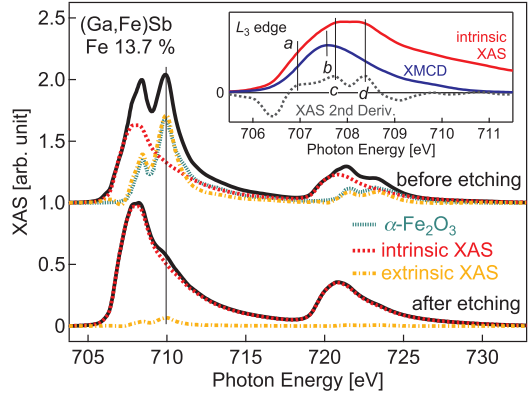}
\caption{XAS and XMCD spectra of the 13.7\% Fe-doped (Ga,Fe)Sb before and after HCl etching.
Intrinsic and extrinsic components are also separately shown by red and orange dashed curves, respectively, together with the spectrum of $\alpha$-Fe$_{2}$O$_{3}$.
Inset shows the intrinsic XAS and XMCD spectra expanded at the Fe $L_{3}$ edge. Second derivative of the XAS spectrum is also shown to emphasize weak features, labeled as $a$-$d$.
Here, the signs of the XMCD and the second derivative spectrum have been reversed.}
\label{XAS}
\end{center}
\end{figure}

In addition to the main peak at $\sim$708 eV, there is a shoulder at $\sim$710 eV, which is more evident in XAS than in XMCD. This can be attributed to Fe$^{3+}$ oxides formed at the surface because the feature at $\sim$710 eV was prominent before etching and disappeared almost completely after etching, as shown in Fig. \ref{XAS}. 
Here, the spectra of both paramagnetic and ferromagnetic samples after etching resemble those of Ge:Fe and (In,Fe)As:Be.
Using these two spectra containing the different degrees of contribution from surface oxides, it was possible to deduce the intrinsic spectra as ${\rm [XAS]_{int}} \propto {\rm [XAS]_{a}} - p{\rm [XAS]_{b}}$. Here, [XAS]$_{\rm a}$ and [XAS]$_{\rm b}$ denote the XAS spectrum after and before etching, respectively, and $p$ was chosen so that the shoulder at $\sim$ 710 eV vanished, or the second derivative of [XAS]$_{\rm int}$ did not show a peak at $\sim$710 eV as shown in the inset of Fig. \ref{XAS}. 
The extrinsic contribution from the surface oxides to the XAS spectra was also extractable in a similar manner as ${\rm [XAS]_{ext}} \propto {\rm [XAS]_{a}} - q{\rm [XAS]_{b}}$, where $q$ was chosen so that [XAS]$_{\rm ext}$ became identical to the spectra of $\alpha$-Fe$_{2}$O$_{3}$ shown by the green dashed curve.
Thus obtained intrinsic and extrinsic components are separately shown in Fig. \ref{XAS} by red and orange dashed curves, respectively, for both spectra before and after etching. 
From this procedure, it was found that the extrinsic contribution to the XAS was almost $\sim$ 60 \% before etching, and was significantly reduced to $\sim$ 4\% after etching, which guarantees the efficiency of HCl etching.

One may suspect that, because the spectra look similar to those of bcc Fe, the magnetism may originate from nanoscale metallic Fe precipitates in the samples. 
In order to rule out this possibility, details of the ``intrinsic XAS" and XMCD spectra at the Fe $L_{3}$ edge are shown in the inset of Fig. \ref{XAS} together with the second derivative XAS spectrum to highlight weak shoulders on the XAS spectrum labeled $a-d$.
Here, the signs of XMCD and the second derivative XAS spectra are reversed for ease of comparison.
The XAS spectrum of (Ga,Fe)Sb mainly consists of two features $c$ and $d$, which do not coincide with the feature in XMCD (labeled $b$).
This is not the case for bcc Fe, where the XAS spectrum consists of a broad single peak, and the peak positions of XAS and XMCD indeed coincide \cite{Sakamoto:2016aa}.
Such a peak-position difference was also found in (In,Fe)As:Be \cite{Sakamoto:2016aa} and Ge:Fe \cite{Wakabayashi:2016aa}, and might be a universal spectral feature of Fe-doped semiconductors. 
Note that there is also a shoulder at $\sim$710 eV in the XMCD spectra originating from Fe$^{3+}$ oxides, but the contribution is much smaller.


\begin{table}
\begin{center}
\caption{Spin and orbital magnetic moments of Fe in (Ga,Fe)Sb in comparison with those of (In,Fe)As:Be, Ge:Fe, and bcc Fe metal. All the values have been estimated using the XMCD sum rules [Eq.(1) and Eq.(2)] except for the ones in the third and the forth rows, which were computed using the GGA and GGA+$U$ methods.}

\begin{tabular}{p{10.5em}p{7em}p{3.5em}p{3.5em}}
\hline\hline
& \hfil $m_{\rm orb}/m_{\rm spin}$ \hfil & \hfil$m_{\rm orb}$\hfil & \hfil$m_{\rm spin}$\hfil \\
\hline
${\rm Ga_{0.94}Fe_{0.06}Sb}^{\ast}$ & \hfil0.13 $\pm$ 0.01\hfil & \hfil0.05$^{\dagger}$\hfil & \hfil0.37$^{\dagger}$\hfil\\
${\rm Ga_{0.863}Fe_{0.137}Sb}^{\ast}$ & \hfil0.13 $\pm$ 0.01\hfil  & \hfil0.14$^{\dagger}$\hfil & \hfil1.07$^{\dagger}$\hfil\\
${\rm In_{0.9}Fe_{0.1}As}$:Be$^{\ast}$ \cite{Sakamoto:2016aa} & \hfil0.10 $\pm$ 0.02\hfil & \hfil0.17$^{\ddagger}$\hfil & \hfil1.75$^{\ddagger}$\hfil\\
${\rm Ge_{0.935}Fe_{0.065}}^{\ast}$ \cite{Wakabayashi:2016aa} & \hfil0.11 $\pm$ 0.03\hfil & \hfil0.14$^{\dagger}$\hfil & \hfil1.29$^{\dagger}$\hfil\\
Fe bcc \cite{Chen:1995aa} & \hfil0.043 $\pm$ 0.001\hfil & \hfil0.085\hfil  & \hfil1.98\ \ \hfil\\
\hline\hline
\multicolumn{4}{l}{$^{\dagger}$Values at $\mu_{0}H$\rm = 1 T and $T$ = 5 K; $^{\ddagger}$Values at 5 T and 20 K}\\
\multicolumn{4}{l}{$^{\ast}$Fe$^{2+}$ configurations was assumed;}\\

\end{tabular}

\label{moment}
\end{center}
\vspace{0cm}
\end{table}

By applying the XMCD sum rules, we have estimated the spin and orbital magnetic moments of Fe\cite{Carra:1993aa,Thole:1992aa} as follows.
\begin{eqnarray}
&\displaystyle m_{\rm orb}=-\frac{4\int_{L_{2,3}}{\rm XMCD}\ {\rm d}\omega}{3\int_{L_{2,3}}{\rm XAS}\ {\rm d}\omega}n_{h},\\
&\displaystyle m_{\rm spin}=-\frac{6\int_{L_{3}}{\rm XMCD}\ {\rm d}\omega-4\int_{L_{2,3}}{\rm XMCD}\ \rm{d}\omega}{\int_{L_{2,3}}{\rm XAS}\ \rm{d}\omega}n_{h},\ \ \ 
\end{eqnarray}
where $m_{\rm orb}$ and $m_{\rm spin}$ are the orbital and spin magnetic moments in units of $\mu_{\rm B}$, respectively, and  $n_{h}$ the number of 3$d$ holes. 
Here, we have ignored the magnetic dipole term, which is negligibly small for an atomic site with high symmetry such as $T_{d}$ or $O_{h}$ \cite{Stohr:1995aa}, and $n_{h}$ was set to 4 assuming the valence of Fe is 2+ with six 3$d$ electrons as implied by the density functional theory calculation \cite{Lin:2017aa}.
The correction factor of 0.875 for Fe$^{2+}$ ion \cite{Teramura:1996aa, Piamonteze:2009aa} was used to estimate the spin magnetic moment.
Note that if we assume the Fe$^{3+}$ state with five 3$d$ electrons and the correction factor to be 0.685, the spin and orbital magnetic moment would be changed by a factor of 1.6 ($=(0.875/0.685)\times(5/4))$.
The raw XAS and XMCD spectra after etching are used for the sake of simplicity because extrinsic contribution was only a few percent.

Table \ref{moment} summarizes the spin and orbital magnetic moments of various Fe compounds including other Fe-doped semiconductors and bcc Fe.
It was found that the $m_{\rm orb}$/$m_{\rm spin}$ ratio of (Ga,Fe)Sb is substantially larger than that of bcc Fe. This may be due to the stronger localization of Fe 3$d$ electrons \cite{stohr1999exploring} in bulk (Ga,Fe)Sb, or at the interface between the Fe-rich and Fe-poor regions, where the translational symmetry is broken \cite{Edmonds:1999aa}. 
The large value of $m_{\rm orb}$/$m_{\rm spin}$  suggests that a considerable fraction of Fe atoms have the valence of 2+ with six 3$d$ electrons because the orbital magnetic moments would be quenched if all the Fe atoms took the high-spin Fe$^{3+}$ (3$d^{5}$) configuration. Nevertheless, there probably exist a significant amount of Fe$^{3+}$ ions as well considering the inhomogeneous nature of the material.
Such a large orbital magnetic moment was also observed for (In,Fe)As:Be \cite{Sakamoto:2016aa} and Ge:Fe \cite{Wakabayashi:2016aa}. This fact and the similar spectral line shapes among the above-mentioned Fe-doped FMSs imply that the local electronic structure of Fe is similar and the ferromagnetism has a common origin.
Note that the $m_{\rm orb}$/$m_{\rm spin}$ ratio is more unambiguous than the absolute values of $m_{\rm orb}$ and $m_{\rm spin}$ since it does contain neither the uncertainty in estimating $n_{h}$ nor extrinsic contributions to the XAS area.

The deduced spin magnetic moments at $H$ = 1T and $T$ = 5 K for both the paramagnetic and ferromagnetic samples are small (0.37 $\mu_{\rm B}$/Fe and 1.07 $\mu_{\rm B}$/Fe, respectively) compared with the ionic value of 4 $\mu_{\rm B}$ \footnote{If the total moment of 4.0 $\mu_{\rm B}$ is assumed, the induced magnetic moment of the paramagnetic sample would become 1.0 $\mu_{\rm B}$ at $T=5$ K and $\mu_{\rm 0}H=1$ T following the Brillouin function.} and the experimental saturated magnetic moment of 2.4-2.9 $\mu_{\rm B}$ measured at the same temperature of 5 K \cite{Tu:2015aa}.
The cause of the small deduced magnetic moment is unclear at this stage, but it can be attributed to the possible existence of Fe$^{2+}$ oxides component in the XAS spectra because the TEY-mode detection of XAS and XMCD signals is surface sensitive with the probing depth of $\sim$3-5 nm.

\subsection{Resonance Photoemission}

\begin{figure*}
\begin{center}
\includegraphics[width=17.0 cm]{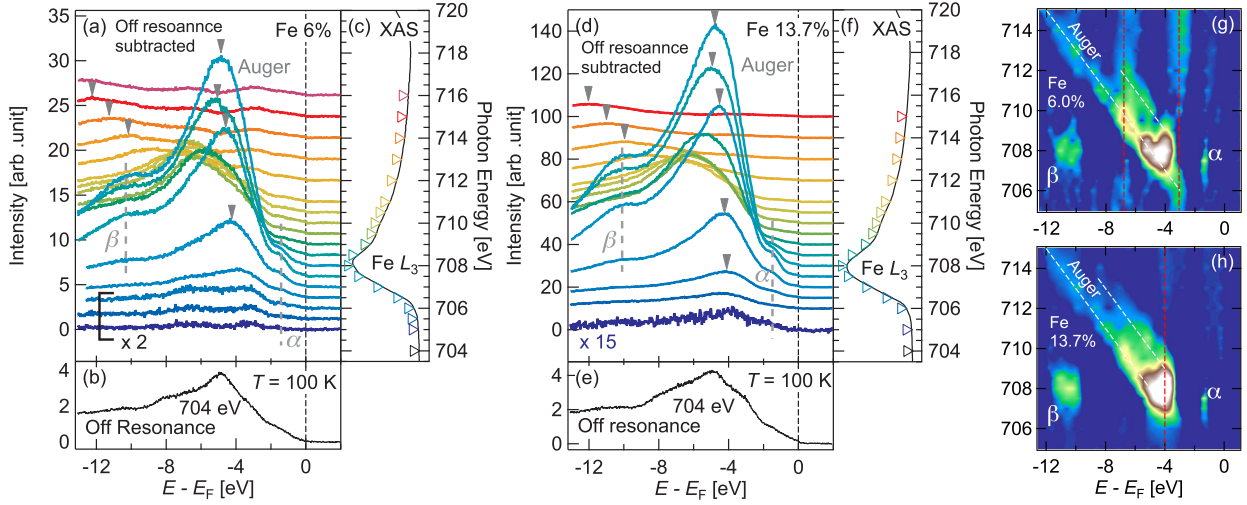}
\caption{(a), (d) Resonance photoemission spectroscopy (RPES) spectra of (Ga,Fe)Sb with 6\% Fe and 13.7\% Fe taken with photon energies across the Fe $L_{3}$ absorption edge. The color and the base line positions of the spectra represent the photon energies as depicted by triangles on the XAS spectra in panels (c) and (f). Here, the off-resonance spectra shown in panels (b) and (e) have been subtracted from all the spectra. (g), (h) False color plots of the second derivatives of the RPES spectra in panels (a) and (d). The dispersive normal Auger peak and the non-dispersive resonance features are indicated by white dashed lines and red solid lines, respectively. }
\label{RPES}
\end{center}
\end{figure*}

Figures \ref{RPES}(a) and \ref{RPES}(d) show RPES spectra of the 6\% and 13.7\% Fe-doped samples, respectively, taken across the Fe $L_{3}$ absorption edge from 705 eV to 716 eV.
The color and the baseline positions represent the photon energies indicated by triangles on the XAS spectra in Figs. \ref{RPES}(c) and \ref{RPES}(f). Here, off-resonance spectra taken with the photon energy of 704 eV shown in Figs. \ref{RPES}(b) and \ref{RPES}(e) have been subtracted to emphasize the resonance behavior. 
Note that the units of the vertical axes of Figs. \ref{RPES}(a), \ref{RPES}(b), \ref{RPES}(d), and \ref{RPES}(e) are the same. 
For both samples with different Fe contents, a strong normal Auger peak dispersing with incident photon energy was observed.
This suggests that the Fe 3$d$ electrons in (Ga,Fe)Sb have itinerant character, as already suggested by the XAS measurements, since Auger decay is a consequence of faster screening of the core hole than the recombination of the excited electron with the core hole and is normally observed in metallic systems \cite{Hufner:2000aa, Levy:2012aa}.
It is worth mentioning that such a strong Auger peak was also observed in the case of Fe-doped Ge \cite{Sakamoto:2017aa}. 
In addition to the Auger peak, there are resonantly enhanced features around -1.7 eV and -10.3 eV, denoted by $\alpha$ and $\beta$, respectively. 
Unlike the Auger peak, these peaks do not disperse with incident photon energy, representing a local excitation of Fe 3$d$ states including charge transfer from ligand orbitals accompanying the photoemission process. 
Such resonance features are widely observed in transition-metal oxides, where the 3$d$ electrons are well localized and strongly correlated \cite{Fujimori:1986aa, Fujimori:1987aa, Lad:1989aa, Kang:2002aa}. 
Those local excitation peaks were almost absent in the case of Fe-doped Ge (Ref. \cite{Sakamoto:2017aa}), suggesting that Fe 3$d$ electrons in (Ga,Fe)Sb are more correlated or more localized than those in Ge:Fe.
Considering that the Fe 3$d$ density of states (DOS) should not be located as deep as 10 eV below $E_{\rm F}$, at least feature $\beta$ can be attributed to a charge-transfer satellite.
In order to highlight subtle features, the second derivative image of the RPES spectra are shown in Figs. \ref{RPES}(g) and \ref{RPES}(h), where dispersive normal Auger peak and resonantly enhanced peaks $\alpha$ and $\beta$ can be seen as described above. 
These features are commonly seen in both samples, however, there exists a difference as indicated by red dashed lines in Figs. \ref{RPES}(g) and \ref{RPES}(h).
For the 6\% Fe sample, two features are observed around 6.8 eV and 3.1 eV in the entire photon energy range, while for the 13.7\% Fe sample, there is only one feature around 4 eV. This may imply the difference in the electronic structure between the two samples. 
Probably the 3$d$ states are more localized or more correlated in the 6\% Fe-doped sample because the distance between adjacent Fe atoms are longer and the Fe 3$d$ band width would be narrower than in the 13.7\% Fe-doped sample.

\begin{figure}
\begin{center}
\includegraphics[width=8.5 cm]{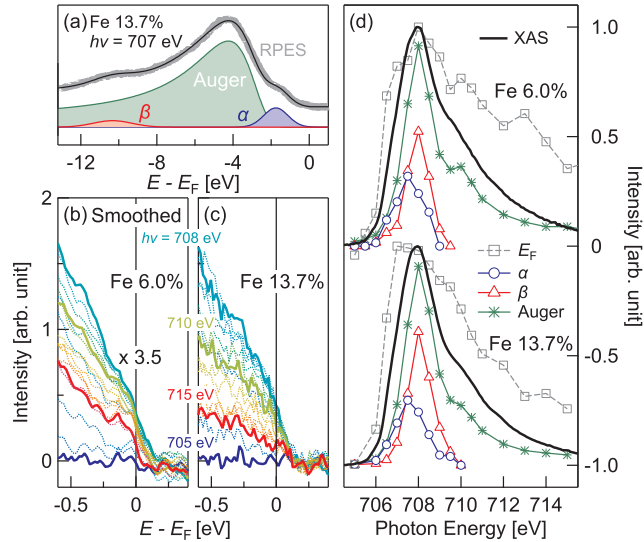}
\caption{(a) RPES spectrum of the 13.7\% Fe-doped sample taken with $h\nu=707$ eV (grey thick curve) and fitting results (black curve). Fitting components of two Gaussian functions for features $\alpha$ and $\beta$ and one asymmetric Gaussian function for Auger peak are also shown separately.
(b), (c) RPES spectra of the 6\% and 13.7\% Fe-doped samples near $E_{\rm F}$. Here, the color is the same as those used in Figs. \ref{RPES}(a) and \ref{RPES}(d). 
Note that the RPES spectra of the 6\% Fe-doped sample have been smoothed and multiplied by 3.5 for easy comparison.
(d) Constant-initial-state (CIS) spectra for features $\alpha$ and $\beta$, and the intensity at $E_{\rm F}$ and the Auger peak intensity as functions of photon energy. For comparison, XAS spectra are also plotted.}
\label{CIS}
\end{center}
\end{figure}

For the purpose of examining the nature of resonance enhancement, it is useful to plot the intensity at a fixed binding energy as a function of incident photon energy, namely, a constant-initial-state (CIS) spectrum. 
In order to eliminate the effect of the overlapping Auger peak from the CIS spectra and to extract the resonance behavior of features $\alpha$ and $\beta$, we have employed curve fitting as shown in Fig. \ref{CIS}(a). 
In the figure, resonance peaks $\alpha$ and $\beta$ are fitted by Gaussian functions and the Auger peak with its tail is fitted by an asymmetric Gaussian function introduced in Ref. \cite{Schmid:2014aa}.
CIS spectra for features $\alpha$ and $\beta$ were obtained from the peak areas of the Gaussian functions. This is an $ad$ $hoc$ procedure, but still it provides a reasonable description of the resonance enhancement.

Figures \ref{CIS}(b) and \ref{CIS}(c) show the RPES spectra on an expanded scale near $E_{\rm F}$. The data of the 6\% Fe-doped sample have been smoothed and multiplied by 3.5 for ease of comparison with the data of the 13.7\% Fe-doped sample.
As can be seen from the figures, the spectral intensities at $E_{\rm F}$ for both samples are enhanced on resonance, suggesting that Fe 3$d$ states have finite contribution to the DOS at $E_{\rm F}$.
Here, the CIS spectrum at $E_{\rm F}$ is defined as the area of RPES spectra between $E-E_{\rm F} = -0.6$ eV and 0.2 eV.


Thus obtained CIS spectra for features $\alpha$, $\beta$, and the intensity at $E_{\rm F}$ are plotted in Fig. \ref{CIS}(d) together with the Auger peak height as a function of photon energy and the XAS spectra.
The CIS spectrum for feature $\alpha$ is peaked at $h\nu=707.5$ eV, but that for feature $\beta$ is peaked at a higher photon energy of $h\nu=708$ eV for both samples. 
The difference in the CIS peak positions implies that the broad XAS spectra actually consist of different kinds of excitations, which may correspond to peaks $c$ and $d$ in the inset of Fig. \ref{XAS}, probably involving different types of 3$d$ orbitals, $i$.$e$., $t_{2}$ and $e$ orbitals, rather than excitation into a single kind of broad metallic bands as in the case of bcc Fe metal.
We note that the CIS spectra of features $\alpha$ and $\beta$ vanish around the photon energy of 710 eV, at which the extrinsic Fe$^{3+}$ shoulder exists on each XAS spectrum. 
Therefore, the observed resonance features are not from surface oxides but are most likely intrinsic.

\section{Discussion}

\begin{figure}
\begin{center}
\includegraphics[width=8.4 cm]{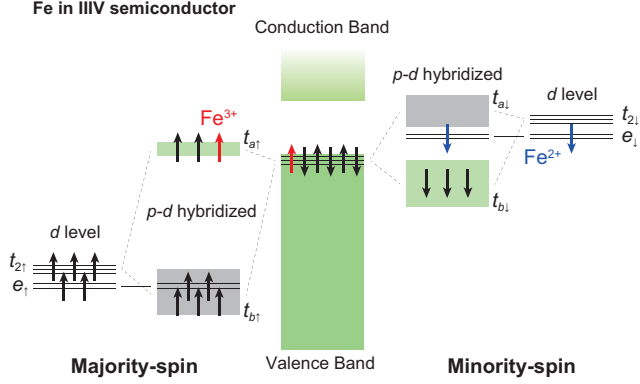}
\caption{Schematic energy diagram of Fe-doped III-V semiconductors. The $d$ levels of Fe are shown on the sides of the figure, and the valence band and the conduction band of the host GaSb are shown by green boxes in the middle. 
Due to $p$-$d$($t_{2}$) hybridization, the $t_{2}$ orbitals are split into the bonding ($t_b$) and anti-bonding ($t_a$) states having both $p$ and $t_{2}$ characters. States with predominant $p$ and $t_{2}$ characters are indicated by green and gray boxes, respectively. The electron occupancy of Fe$^{3+}$ state is illustrated by red and black arrows while that of Fe$^{2+}$ by blue and black arrows.}
\label{diagram}
\end{center}
\end{figure}

Figure \ref{diagram} illustrates the basic electronic structure of Fe in GaSb. The valence and conduction bands of GaSb are shown in the middle by large green boxes, and the Fe 3$d$ levels are shown on both right-hand and left-hand sides. Each arrow represents an electron with a specific spin.
At the substitutional sites, the Fe 3$d$ level is split into the doubly-degenerate lower $e$ level and the triply-degenerate higher $t_{2}$ level due to the crystal field of $T_{d}$ symmetry.
Due to the strong $p$-$d$($t_{2}$) hybridization, the Fe $t_{2}$ orbitals and the ligand Sb $p$ orbitals form bonding ($t_{b}$) and anti-bonding ($t_{a}$) states with mixed $t_{2}$ and $p$ characters.
Those levels with predominant $t_{2}$ and $p$ characters are indicated by gray and green boxes, respectively.
As for the majority-spin states, where the $t_{2\uparrow}$ level is located well below the valence-band maximum (VBM) because of the relatively high Sb 5$p$-derived valence-band position \cite{Vurgaftman:2001aa}, the bonding states have predominant Fe $t_{2}$ character, and the anti-bonding states predominant Sb $p$ character. 
On the other hand, the minority-spin bonding states ($t_{2\downarrow}$) consist primarily of Sb $p$ orbitals, and the anti-bonding states of Fe $t_{2}$ orbitals. 

If Fe takes the valence of 3+ with five 3$d$ electrons, the $t_{a\uparrow}$ level is fully occupied and there exists no minority-spin 3$d$ electrons (except for the small contribution in the $t_{b\downarrow}$ states). 
On the other hand, if the valence of Fe is 2+, the sixth 3$d$ electron occupies the $e_{\downarrow}$ level and one hole resides in the $t_{a\uparrow}$ level. The $e_{\downarrow}$ level would be at the Fermi level since it is doubly degenerate and half occupied.
In Fig. \ref{diagram}, the Fe$^{3+}$ electronic structure is represented by red and black arrows, while Fe$^{2+}$ by blue and black arrows.

From the XMCD measurements, it was suggested that there is a considerable $3d^{6}$ contribution to the ground state of Fe as in (In,Fe)As:Be \cite{Kobayashi:2014aa, Sakamoto:2016aa} and Ge:Fe \cite{Wakabayashi:2016aa}. 
This suggests that there may exist the long-range $p$-$d$ exchange interaction mediated by holes in $t_{a\uparrow}$ states as in the case of (Ga,Mn)As.
Considering the coexistence of Fe$^{2+}$ and Fe$^{3+}$ states, the short-range double exchange interaction within the $e_{\downarrow}$ orbitals would also be present.
Note that the finite Fe PDOS at the Fermi level found by the RPES measurements can be attributed to either $t_{a,\uparrow}$ or $e_{\downarrow}$ states in this scenario.
If this is the case, the double exchange interaction would be more important than the $p$-$d$ exchange interaction considering that the ferromagnetism with similar $T_{\rm C}$ is also observed in the $n$-type (In,Fe)As and (In,Fe)Sb, where the $s$-$d$ exchange interaction is significantly weaker than the $p$-$d$ exchange interaction.
In real (Ga,Fe)Sb samples, it was reported that there is Fe concentration fluctuation especially when Fe is heavily doped \cite{Tu:2015aa, Tu:2016aa}. 
The double exchange interaction would be locally strong in Fe-rich regions and stabilize the local ferromagnetic order.
This may explain the observed convex line shape of $M$-$T$ curves \cite{Tu:2015aa} indicating the existence of superparamagnetism even above $T_{\rm C}$. 

The Fe$^{2+}$ scenario described above, however, may not explain the fact that the carrier concentration of (Ga,Fe)Sb obtained by Hall measurements is not more than $\sim10^{19}$ cm$^{-3}$ (Ref. \cite{Tu:2015aa}), two orders of magnitude smaller than doped Fe atoms. 
One possible explanation is that the carriers are strongly trapped inside the Fe-rich regions and macroscopic carrier transport occurs via hopping between those Fe-rich regions. Such a model was introduced by Kaminski and Das Sarma \cite{Kaminski:2003aa} and applied to Ge:Mn \cite{Pinto:2005aa}, Ge:Fe \cite{Ban:2017aa}, and (Zn,Cr)Te \cite{Sreenivasan:2007aa} to describe their insulating/semiconducting natures and low carrier concentrations ($\sim$10$^{18}$ cm$^{-3}$ for Ge:Mn \cite{Pinto:2005aa} and Ge:Fe \cite{Ban:2014aa},  $\sim$10$^{15}$ cm$^{-3}$ for (Zn,Cr)Te \cite{Saito:2002aa}).
Note that, although the low carrier concentration of (Ga,Fe)Sb would be explained by Fe$^{3+}$ scenario instead, where only the $p$-$d$ exchange interaction is present, it seems difficult to explain why ferromagnetism is universally observed in the other Fe-doped FMSs regardless of the carrier type. It is worth mentioning that a recent theoretical calculation \cite{Shinya:2018aa} has pointed out the important role of superexchange interaction.
In order to resolve the puzzle and to fully understand the peculiar nature of the ferromagnetism in the Fe-doped FMSs, further theoretical and experimental studies are necessary.

\section{Summary}
In the present work, we have studied the electronic structure and the magnetism of the novel $p$-type ferromagnetic semiconductor (Ga,Fe)Sb, whose Curie temperature exceeds room temperature, using XAS, XMCD, and RPES.
The line shapes of XAS and XMCD spectra suggested that the ferromagnetism is of intrinsic origin. 
XMCD sum rules yielded an unquenched large orbital moment, implying the 3$d^{6}$ configuration of Fe. 
The valence-band RPES spectra showed a dispersive Auger peak and non-dispersive resonantly enhanced peaks, which are fingerprints of the itinerant and correlated nature of the Fe 3$d$ electrons, respectively.
It was also found that there is a finite Fe PDOS at the Fermi level. This has been attributed to majority-spin antibonding $p$-$d$($t_{2}$) hybridized state and/or minority spin $e$ state, both of which can play a role in stabilizing the ferromagnetic order through $p$-$d$ exchange interaction and double exchange interaction, respectively.

\section*{Acknowledgments}

This work was supported by Grants-in-Aid for Scientific Research from the JSPS (Grants No. 15H02109, No. 15K17696, and No.16H02095), and CREST of JST (Grant No. JPMJCR1777).
The experiment was done under the Shared Use Program of JAEA Facilities (Proposal No. 2015B-E24 and No. 2016A-E27) with the approval of the Nanotechnology Platform Project supported by MEXT.
The synchrotron radiation experiments were performed at the JAEA beamline BL23SU in SPring-8 (Proposal No. 2015B3881 and No. 2016A3831).
S.S., L.D.A., Y.K.W., and M.H. acknowledge support from the Program for Leading Graduate Schools; L.D.A., Y.K.W., and M.H. acknowledge support from the JSPS Fellowship for Young Scientists. 
P.N.H. acknowledges support from the Yazaki Memorial Foundation for Science and Technology, the Murata Science Foundation, and the Toray Science Foundation.
A.F. is an adjunct member of Center for Spintronics Research Network (CSRN), the University of Tokyo, under Spintronics Research Network of Japan (Spin-RNJ).

\bibliography{BibTex_all_GaFeSb.bib}

\end{document}